\documentclass[aps,prb,twocolumn,showpacs,superscriptaddress]{revtex4-1}
\usepackage{amsfonts,amsmath,amssymb}
\usepackage[colorlinks=true,linkcolor=blue,pagecolor=blue,filecolor=blue,menucolor=blue,urlcolor=blue,citecolor=blue,anchorcolor=blue]{hyperref}%
\usepackage[dvips]{graphics}
\usepackage{graphicx}
\usepackage{bm}
\usepackage{color}

\begin{document}

\title{Thermospin effects in superconducting heterostructures}
\author{I. V. Bobkova}
\affiliation{Institute of Solid State Physics, Chernogolovka, Moscow
  reg., 142432 Russia}
\affiliation{Moscow Institute of Physics and Technology, Dolgoprudny, 141700 Russia}
\author{A. M. Bobkov}
\affiliation{Institute of Solid State Physics, Chernogolovka, Moscow reg., 142432 Russia}

\date{\today}

\begin{abstract}
Recently thermally created pure spin currents were predicted for Zeeman-split superconductor/normal metal heterostructures. Here it is shown that this "thermospin" current can lead to an accumulation of a pure spin imbalance in a system. The thermally induced spin imbalance can reach the value of Zeeman splitting of the superconducting DOS and strongly influences superconductivity in the heterostructure. Depending on the temperature difference between the superconductor and the normal reservoir it can enhance the critical temperature of the superconductor or additionally suppress the zero-temperature superconducting state. The last possibility gives rise to an unusual superconducting state, which starts to develop at finite temperature.
\end{abstract}

\maketitle

\section{Introduction}

The coupling of superconductivity and magnetism in hybrid structures is a focus of the research in condensed matter physics now. It leads to many interesting fundamental effects. Among them are Cooper pairs with finite center of mass momentum, triplet odd frequency Cooper pairs \cite{eschrig15,buzdin05,bergeret05,keizer06,alidoust10,robinson10,khaire10,anwar10,bergeret14,gomperud15,jacobsen15_1,jacobsen15_2,alidoust15}, longer spin lifetimes and spin charge separation \cite{yang10,hubler12,quay13,wolf13,wakamura14,wolf14,bobkova15,silaev15,krishtop15,bobkova16,virtanen15,quay16} and fully spin-polarized electric currents \cite{giazotto08}. The largest part of the mentioned effects is based on the coupling of charge and spin degrees of freedom and, therefore, is of interest for superconducting spintronics \cite{linder15}.

Recently it was also shown that coupling of heat transport and charge degrees of freedom in superconductor/ferromagnet (S/F) hybrid structures also results in quite interesting effects. The so-called giant thermoelectric effect was predicted in S/F heterostructures with applied in-plane magnetic field \cite{machon13,ozaeta14} and in superconductors with magnetic impurities \cite{kalenkov12}. The theoretical prediction of the giant thermoelectric effect in S/F heterostructures was experimentally realized in Refs.~\onlinecite{kolenda16,kolenda16_1}. It was also predicted that the use of doubly Zeeman-split superconducting bilayers enhances the thermoelectric effect significantly \cite{linder16}. Further the thermoelectric effect was proposed in superconducting hybrids with spin-textured materials without the need to apply an external magnetic field \cite{bathen16}.

The coupling of heat and spin degrees of freedom is also a hot topic now. In particular, heat transport with
participation of spin-triplet Cooper pairs was discussed \cite{silaev2017}. It was reported that in addition to thermally created electric currents in Zeeman-split superconducting systems there should exist also thermally created spin currents \cite{ozaeta14,linder16}. In contrast to thermoelectric currents, these spin currents do not require a spin polarization of the barrier between the materials of the hybrid and, therefore, can exist even in superconductor/normal metal hybrids provided that there is a spin-dependent particle-hole asymmetry on at least one side of the interface. It is worth noting that this spin Seebeck effect (it can be also called by "thermospin" effect) is also known for normal (nonsuperconducting) systems \cite{slachter10,bauer12}, but the great advantage of using  superconductors is that the spin-dependent particle-hole asymmetry here is very large due to the presence of the superconducting gap at the Fermi level. For this reason the thermally induced electric and spin currents in superconducting structures are "giant" with respect to their analogs in normal systems. The thermally induced pure spin currents were  predicted at S/N interfaces with Zeeman-split superconductors \cite{ozaeta14}, at interfaces between two Zeeman-split superconductors \cite{linder16} and also in more complicated systems, composed of two Josephson junction interlayers \cite{linder16} and Josephson junctions with spin-textured materials \cite{bathen16}, where the additional superconducting phase control of the spin current can be achieved.

In this paper we show that the existence of such a "thermospin" effect is a natural source of thermally activated pure spin imbalance in superconductors. We explore the maximal value of this imbalance and its influence on the superconductivity of S/N bilayer. We predict that in S/N heterostructures with a temperature difference between N and S parts the spin imbalance can result in the enhancement of its critical temperature. A more interesting case is when it leads to the enhancement or even appearance of superconductivity upon heating the sample. This effect can be a clear signature of the presence of the thermally induced spin imbalance in the superconductor.

There are other ways to create spin imbalance in hybrid structures. One of them is the injection of an electric current from a strong ferromagnet \cite{johnson85}. It is inevitably accompanied by electric currents or voltages, that is, it leads to coupling charge, spin and heat degrees of freedom simultaneously. Another way is photoassisted spin imbalance \cite{virtanen15}. But in this case the spin imbalance generation mechanism is strongly different. There is no an external source of spins in the system. The spin imbalance is acquired here not due to spatial movement of the spin but due to internal redistribution of the quasiparticles between the Zeeman-split energy subbands. This spin imbalance is absent in the absence of the spin-flip processes.

The paper is organized as follows. In Sec.~\ref{spin_imb} we calculate the thermally induced spin imbalance, discuss underlying qualitative physics and describe the possible setups, where the effect can be observed. In Sec.~\ref{OP} we calculate and discuss the influence of this spin imbalance on the superconductivity in the system. Sec.~\ref{spin_flip} is devoted to the influence of the elastic spin-flip processes on the predicted effects and discussion of the most favorable for the experimental realization regimes of parameters. Our conclusions are presented in Sec.~\ref{conclusions}.

\section{spin imbalance}

\label{spin_imb}

\begin{figure}[!tbh]
  \centerline{\includegraphics[clip=true,width=3.0in]{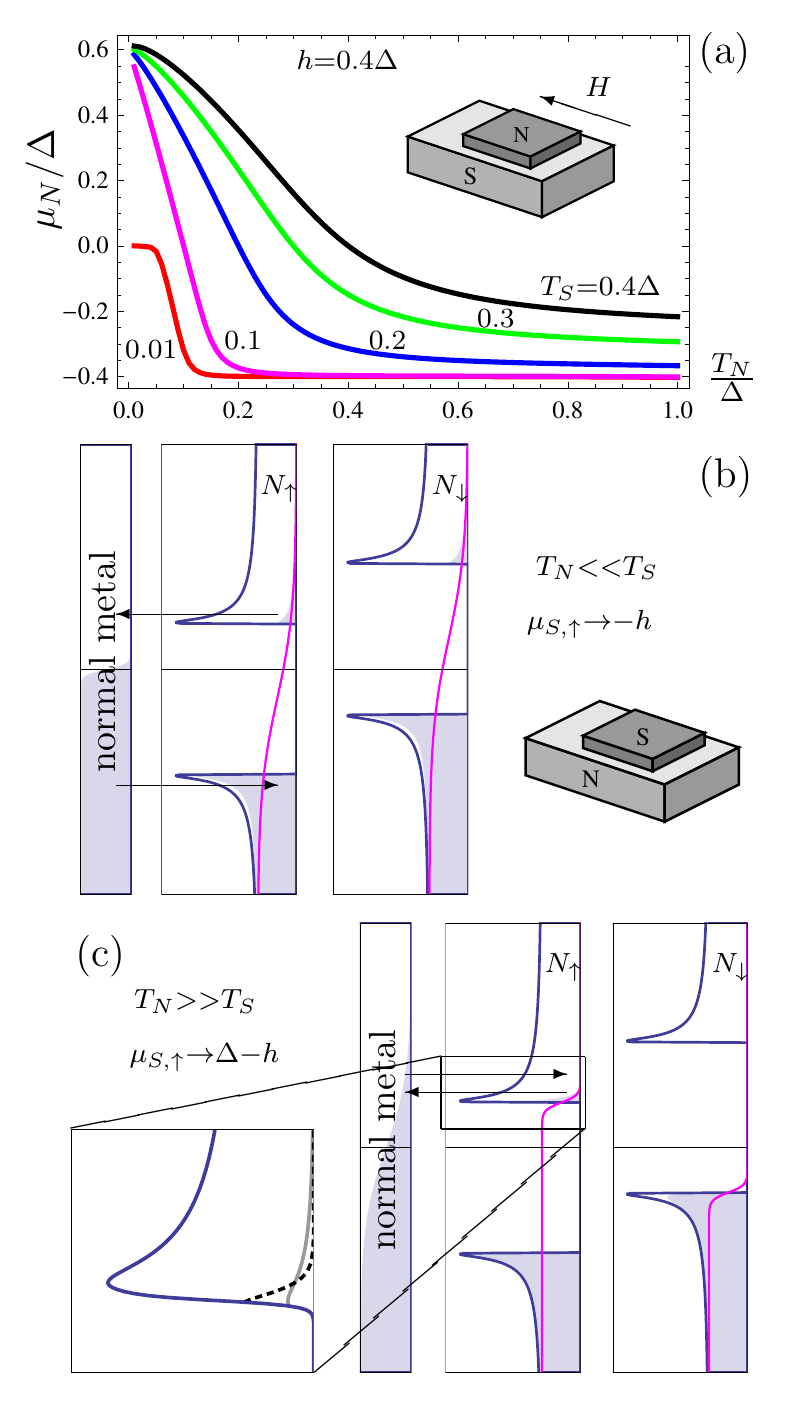}}
   \caption{(a) Spin imbalance, accumulated in the normal metal as a function of $T_N$. Different curves correspond to different temperatures of the superconducting reservoir $T_S$. (b) and (c) Schematic illustration of the thermospin effect for the case when the reservoir is normal. The DOS in the N layer (left) and in the S layer (right) are shown as functions of $\varepsilon$ (vertical axis). The shaded parts are occupied by electrons. The pink lines are the distribution function in S. The arrows correspond to the directions of the electron flow between the spin-up subband and the N layer. The left part of panel (c) represents a marked region of the right figure on a larger scale. The blue line is a superconducting DOS profile, the grey line represents the thermal quasiparticle filling corresponding to $T_N$ and the black dashed line shows the actual quasiparticle filling in the superconductor at $T=T_S$.}
\label{imbalance}
\end{figure}

At first we describe qualitatively the mechanism of thermally induced pure spin current generated at the tunnel interface between the normal metal and the Zeeman-split superconductor. Let us consider the corresponding interface, as depicted in Figs.~\ref{imbalance}(a) and (b). The external magnetic field is applied parallel to the thin superconducting film in order to avoid large orbital depairing of superconductivity. In this situation the main effect of the field on the superconductivity is the Zeeman splitting of the spin subbands. The gapped superconducting DOS is spin-split and the DOS for spin-up and spin-down spin subbands are shifted by the value of the Zeeman field $h=\mu_B H$ in the opposite directions from the the Fermi level, as it is shown in panels (b) and (c) of Fig.~\ref{imbalance}. At finite temperature the spin-up DOS above the gap is partially occupied by quasiparticles and the spin-down DOS is partially empty below the gap. If the superconductor temperature $T_S$ differs from the normal metal temperature $T_N$, it leads to the appearance of the quasiparticle currents between each of the subbands and the normal metal. These spin-up and spin-down quasiparticle currents flow in the opposite directions and the electric currents, carried by these flows cancel each other exactly due to the overall particle-hole symmetry of the superconducting DOS. Therefore, there is no thermoelectric effect in this situation \cite{note1}, but there is thermally induced pure spin current.
The underlying physical reason is that the spin-dependent particle-hole asymmetry is very large here for the energies of the order of $\Delta$ as compared to the nonsuperconducting case \cite{bauer12}.

\subsection{Model and basic assumptions}

Let us assume that one of the layers (S or N) serves as a spin reservoir, and the other layer is confined, that is the thermally injected spin is not removed from it. If the thickness of this confined layer is less than the spin relaxation length, the spin imbalance is accumulated in it due to the thermospin effect.
Now we discuss the conditions required for the experimental realization of our proposal. We concentrate on the case when the confined layer is superconducting because it is interesting in the context of the spin imbalance influence on superconductivity.

\begin{figure}[!tbh]
  \centerline{\includegraphics[clip=true,width=3.4in]{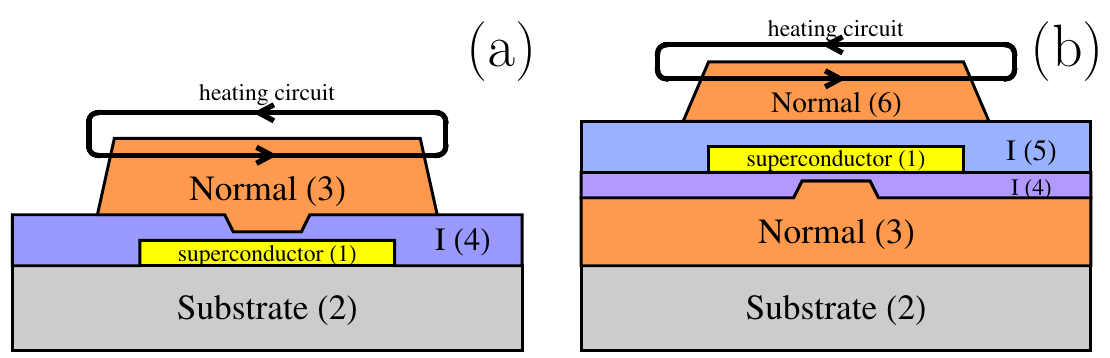}}
   \caption{Sketch of the possible setups, which allow for experimental observation of the thermally induced spin imbalance in the superconductor. Panel (a) corresponds to the case of "cold" superconductor and "hot" normal reservoir, while panel (b) illustrates the opposite case of "hot" superconductor and "cold" normal reservoir.}
\label{setups}
\end{figure}

The temperature difference, which is required for observation of the considered effects, can be obtained, for example, in the setups, sketched in Figs.~\ref{setups}(a) and (b). In both cases the confined S layer [(1) in Figs.~\ref{setups}(a) and (b)] is a thin film. The magnetic field is applied in the plane of this film. Insulating substrate (2) is assumed to be the coldest part of the system and the low temperature of the substrate is maintained by the external conditions.

Let us consider the  case when the normal reservoir is more heated than the confined superconducting layer [Fig.~\ref{setups}(a)]. In this case we assume that the confined layer has a good heat contact with a substrate. The heat contact between the normal reservoir (3) and the confined layer is unavoidable, but is assumed to be weaker than the heat contact between the confined layer and the substrate. To meet this requirement the insulating layer (4) is inserted. In its thinnest part it works as a tunnel barrier forming the S/N interface. At the same time the normal reservoir (3) should be decoupled from the environment as much as possible. Then it can be heated in a controllable way by the Joule heat.

The other case (when the confined layer is at higher temperature) can be realized by inverting the setup described above [Fig.~\ref{setups}(b)]. That is the normal reservoir (3) should be in a good heat contact with the substrate. In this case to heat the confined superconductor (1) one needs an additional resistive element (6). The insulating layer (4) again serves as a barrier between S and N reservoir. Another insulating layer (5) is used for preventing the electric contact between the superconductor and the heating element.

The particular value of the spin imbalance depends on the parameters of the system. Further we calculate this thermally induced spin imbalance under the assumption $\tau_\varepsilon \ll \tau_G \ll \tau_{sf}$, where $\tau_\varepsilon$ is the characteristic energy relaxation time in the layer, $\tau_G$ is the characteristic time, that a quasiparticle spends in it and $\tau_{sf}$ is the spin relaxation time. Under these conditions quasiparticles in the confined layer can be described by a definite electronic temperature $T^e$ (due to the fact that the energy relaxation is the fastest process) and the different spin species have their own chemical potentials due to the weakness of the spin-flip processes. In this case the thermally induced spin flow across the interface is compensated by the counterflow due to the different chemical potentials for spin-up and spin-down quasiparticles and the spin relaxation in the layer can be disregarded because of the inequality $\tau_G \ll \tau_{sf}$. Therefore, the distribution function takes the form of the shifted Fermi function $f_{\uparrow(\downarrow)}=f(\varepsilon-\mu_{\uparrow(\downarrow)},T^e)$, where $\mu_{\uparrow(\downarrow)}=\pm \mu$, and $\mu$ is the resulting spin imbalance. It automatically leads to zero charge and spin flows across the interface, but the heat flow is nonzero due to the finite temperature difference at the interface. Consequently, we should also take into account the heat balance equation, which describes the removal of this heat into the phonon subsystem. The corresponding equation determines the effective temperature of the electron gas in the confined layer, but for strong enough energy relaxation this temperature is very close to the phonon temperature in the layer $T^e \approx T^{ph} = T$. The heat balance equation is discussed in detail in subsection \ref{heat_balance}.

\subsection{Spin imbalance in the normal layer}

Let us first analyze the simplest case when the confined layer, where the spin imbalance is accumulated, is normal. Under our assumptions $\mu_N$ can be found from the condition of zero total spin current through the interface:
\begin{eqnarray}
\int \limits_{-\infty}^\infty d\varepsilon \sum \limits_\sigma \sigma N_\sigma \Biggl[  \tanh \frac{\varepsilon-\sigma \mu_N}{2T_N}-\tanh \frac{\varepsilon}{2T_S}  \Biggr]=0,
\label{spin_imbalance}
\end{eqnarray}
where $\sigma=\uparrow,\downarrow$ in the subscripts and $\pm$ in the expressions, and $N_\sigma$ is the spin resolved DOS in the superconductor.

The spin imbalance $\mu_{N}$, accumulated due to the thermospin effect in the normal layer, is plotted in Fig.~\ref{imbalance}(a) as a function of $T_N$ for different temperatures of the superconductor $T_S$. There are three important features in the behavior of the spin imbalance: (i) it is obvious that $\mu_{N}=0$ at $T_N=T_S$ - there is no thermospin effect at all; (ii) $\mu_{N} \to \Delta-h$ at $T_N \ll T_S$ and (iii) $\mu_{N} \to -h$ for the opposite limit $T_S \ll T_N$. These characteristic features determine the maximal value of the thermally induced spin imbalance in Zeeman-split S/N heterostructures. These limits are violated when $h$ becomes very small (of the order of the lowest temperature).

It is interesting to trace how the thermospin imbalance influences the overall amount of the electron spin accumulated in the normal layer. Without the thermospin imbalance the accumulated spin $S =N_F h$, where $N_F$ is the energy-independent DOS in the normal layer. If we take the spin imbalance into account, then $S =N_F (h+\mu_N)$. Therefore, at $T_N \ll T_S$ we obtain $S \to N_F \Delta {\rm sgn} h$. That is, the spin susceptibility in the normal metal becomes highly nonlinear: it is very high at small $h \lesssim T_N$ and nearly zero for larger $h$ until the superconductivity in the reservoir is destroyed by the Zeeman field. In the opposite case $T_S \ll T_N$ the accumulated spin $S \approx 0$ for a wide range of $h$. This behavior of the spin susceptibility in the normal metal resembles the behavior in a superconductor.

\subsection{Spin imbalance in the superconductor}

The opposite case of spin accumulation in the superconductor $\mu_{S}$ is presented in Figs.~\ref{spin_Delta}(b), \ref{spin_Delta_suppl}(b) and \ref{Delta_orb}(b). This case is more complicated because of the fact that the superconducting order parameter $\Delta$ is very sensitive to the value of the spin imbalance and the particular temperature of the superconductor. At each point of the curves $\mu_S$ and $\Delta$ are determined self-consistently from the following equation, analogous to Eq.~(\ref{spin_imbalance}):
\begin{eqnarray}
\int \limits_{-\infty}^\infty d\varepsilon \sum \limits_\sigma \sigma N_\sigma \Biggl[  \tanh \frac{\varepsilon}{2T_N}-\tanh \frac{\varepsilon - \sigma \mu_S}{2T_S}  \Biggr]=0
\label{spin_imbalance_S}
\end{eqnarray}
and the self-consistency equation for the order parameter $\Delta$ [Eq.~(\ref{selfcons})]. Here the limiting values of the spin imbalance are the same: $\mu_{S} \to \Delta-h$ at $T_S \ll T_N$ and  $\mu_{S} \to -h$ for the opposite limit $T_N \ll T_S$, but  now $\Delta$ is calculated for the given $h$, $T_S$, $T_N$ and $\mu_S$. It is also obvious that $\mu_S \to 0$ at $\Delta \to 0$ or $h \to 0$ independent of $T_N$ and $T_S$ because of zero spin-dependent particle-hole asymmetry in these cases.

Physically it is clear why the thermally induced spin imbalance has just these limiting values. If one neglects the spin relaxation, $\mu_S$ is determined by Eq.~(\ref{spin_imbalance_S}), which in fact means that the spin flow is zero for each of the spin subbands separately. If $T_S \ll T_N$, in the spin-up subband quasiparticles are accumulated at $\varepsilon > \Delta-h$ [see Fig.~\ref{imbalance}(c)]. Under the condition of fast thermalization processes the presence of a large number of quasiparticles at $\varepsilon \sim \Delta-h$ and low temperature of the superconductor just mean that the chemical potential $\mu_\uparrow \sim \Delta-h$ for this spin subband. For the opposite case $T_N \ll T_S$ [see Fig.~\ref{imbalance}(b)] the spin balance for the spin-up subband requires that the spin flow from S to N at $\varepsilon > \Delta-h$ equals to the opposite spin flow from N to S at $\varepsilon < -\Delta-h$. It is possible if the number of quasiparticles at $\varepsilon > \Delta-h$ and $\varepsilon < -\Delta-h$ are equal, that is the distribution function should be symmetric with respect to the gap. Consequently, $\mu_\uparrow = -h$.

\subsection{Heat balance equation}
\label{heat_balance}

The consideration of the previous two subsections was based on the condition of zero electric and spin currents through the interface. But the heat flow is nonzero. In this subsection we discuss the details of the heat balance.

For definiteness we consider the case when the confined layer is superconducting. If we neglect the elastic spin-flip processes completely, that is we assume $\tau_{sf} \to \infty$, from the Keldysh part of the Usadel equation we obtain the following equation for the distribution function $\varphi_\sigma$ \cite{bobkova16}:
\begin{equation}
D\kappa_\sigma \partial_x^2 \varphi_\sigma -I_{\sigma, e-ph} - I_{\sigma, e-e}=0
\label{S5}
\enspace ,
\end{equation}
where the $x$-axis is along the normal to the S/N interface, $\kappa_\sigma=1+|\cosh \theta_\sigma|^2-|\sinh \theta_\sigma|^2$ accounts for the renormalization of the diffusion constant by superconductivity. It is expressed via the retarded normal and anomalous Green's functions
   in terms of the $\theta$-parametrization \cite{belzig99}. The collision integrals $I_{\sigma, e-ph}$ and $I_{\sigma, e-e}$ describe the electron-phonon and electron-electron relaxation processes, respectively. Please note that in the problem under consideration we only consider the so-called "symmetric" types of the quasiparticle nonequilibrium, when the electron and hole components of the distribution function are the same, the distribution function has no $\tau_3$ contribution. This is because the source of the nonequilibrium here is the temperature gradient, which is just of this symmetric type.

We assume that the film thickness in the $x$ direction is smaller than the characteristic relaxation length of the distribution function. Then $\varphi_{\uparrow ,\downarrow}$ does not depend on $x$. Integrating Eq.~(\ref{S5}) over the width $d$ of the film along the $x$-direction and taking into account the boundary conditions for the distribution function \cite{bobkova16}
\begin{equation}
\kappa_\sigma \partial_x \varphi_\sigma \Biggr |_{x=0}=\frac{2G}{\sigma_S}Re[\cosh \theta_\sigma]\biggl( \varphi_\sigma - \tanh \frac{\varepsilon}{2T_N} \biggr)
\label{S6}
\enspace ,
\end{equation}
where $\sigma_S$ is the normal state conductivity of the superconductor and $G$ is the interface conductance, one can obtain that $\varphi_\sigma$ obeys the following equation:
\begin{equation}
-\frac{2Re[\cosh \theta_\sigma]}{\tau_G}\biggl( \varphi_\sigma - \tanh \frac{\varepsilon}{2T_N} \biggr)=I_{\sigma, e-ph} + I_{\sigma, e-e}
\label{S7}
\enspace ,
\end{equation}
where we introduce the characteristic time, that a quasiparticle spends in the superconducting layer $\tau_G=(\sigma_S d)/(GD)$.

Multiplying the kinetic equation Eq.~(\ref{S7}) by the quasiparticle energy $\varepsilon$ and integrating it over energy, one can obtain the heat balance equation, which takes the form:
\begin{gather}
-\frac{2}{\tau_G}\int \limits_{-\infty}^\infty \varepsilon d \varepsilon Re[\cosh \theta_\sigma] \biggl( \varphi_\sigma - \tanh \frac{\varepsilon}{2T_N} \biggr)= \notag \\
\int \limits_{-\infty}^\infty \varepsilon d \varepsilon I_{\sigma, e-ph}
\label{S8}
\enspace ,
\end{gather}
the contribution of the electron-electron relaxation term equals zero because this term conserves the total energy. We assume that the electron-electron relaxation is the fastest relaxation process in the system. Under this condition the distribution function takes the form of the Fermi distribution $\varphi_\sigma = \tanh (\varepsilon-\mu_\sigma)/2T_{S,\sigma}^e$. In each of the spin-subbands this Fermi distribution is characterized by its own chemical potential $\mu_\sigma$ and, in principle, its own effective electronic temperature $T_{S,\sigma}^e$.  Eq.~(\ref{S8}) together with Eq.~(\ref{spin_imbalance_S}) determines the spin imbalance $\mu_\sigma = \pm \mu_S $ and the effective temperature $T_{S,\sigma}^e$ of the electron subsystem of the superconductor for a spin $\sigma$. However, due to the overall particle-hole symmetry $Re[\cosh \theta_\sigma (\varepsilon)]=Re[\cosh \theta_{-\sigma} (-\varepsilon)]$, $Re[\sinh \theta_\sigma (\varepsilon)]=-Re[\sinh \theta_{-\sigma} (-\varepsilon)]$ and $\varphi_\sigma (-\varepsilon)=-\varphi_{-\sigma}(-\varepsilon)$ (the last equality is only valid for the symmetric type of nonequilibrium) the electron-phonon integral \cite{kopnin,virtanen15,bobkova16}
\begin{gather}
I_{\sigma, e-ph} = 8 g_{ph} \int \limits_{-\infty}^\infty d \varepsilon' (\varepsilon'-\varepsilon)^2 {\rm sgn}(\varepsilon'-\varepsilon)\times \nonumber \\
\Bigl[ Re[\cosh \theta_\sigma(\varepsilon)]Re[\cosh \theta_\sigma(\varepsilon')]- \notag \\
Re[\sinh \theta_\sigma(\varepsilon)]Re[\sinh \theta_\sigma(\varepsilon')] \Bigr]\times \notag \\
\Bigl\{ \coth \frac{\varepsilon'-\varepsilon}{2T_S}[\varphi_\sigma (\varepsilon)-\varphi_\sigma (\varepsilon')]-\varphi_\sigma (\varepsilon)\varphi_\sigma (\varepsilon')+1 \Bigr\}
\label{S9}
\end{gather}
also manifests the same symmetry $I_{\sigma, e-ph}(-\varepsilon)=-I_{-\sigma, e-ph}(\varepsilon)$. In this case Eq.~(\ref{S8}) is valid at the same temperature $T_{S,\uparrow}^e=T_{S,\downarrow}^e=T_S^e$ for the both spin directions.

\begin{figure}[!tbh]
  \centerline{\includegraphics[clip=true,width=2.7in]{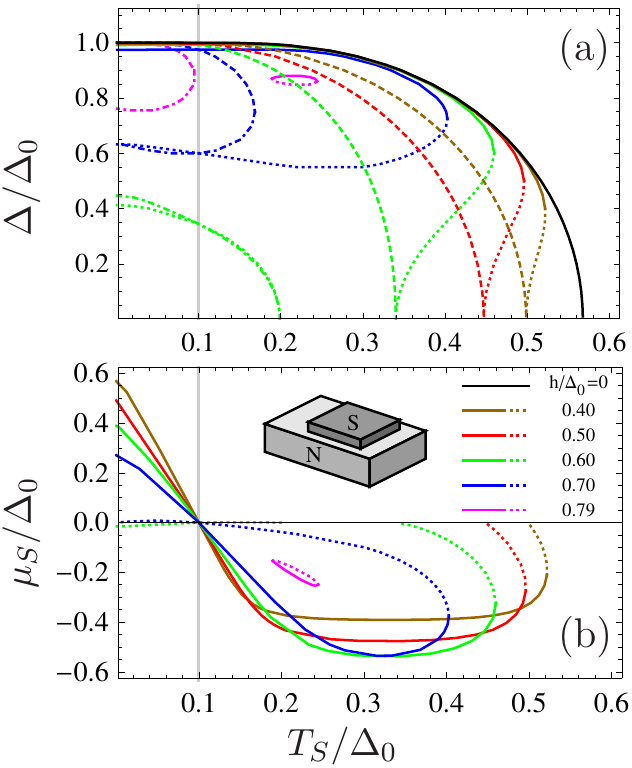}}
   \caption{(a) Superconducting order parameter for $T_N=0.1 \Delta_0$ (vertical gray line) as a function of $T_S$. Here $\Delta_0 \equiv \Delta(T_S=T_N=0,h=0)$. Different colors correspond to different magnetic fields. The dashed lines are results of the order parameter calculation at $\mu_S=0$ (the dashed-dotted parts of the curves represent the absolutely unstable branches of the solutions). The solid lines are the results taking into account the real value of the spin imbalance. The dotted parts of these curves correspond to absolutely unstable branches of the solution. (b) The appropriate spin imbalance $\mu_S$ as a function of $T_S$.}
\label{spin_Delta}
\end{figure}

If the electron-phonon relaxation is strong, that is $\tau_{e-ph} \ll \tau_G$, it is seen from Eq.~(\ref{S9}) that the electron temperature $T_{S}^e$ is very close to the phonon temperature $T_S$. It is this case that is basically considered in our paper. For this reason we omit the superscript $e$ in all the calculations below and in Figs.~\ref{imbalance}, \ref{spin_Delta}-\ref{Delta_orb}. But in general we work with electronic temperatures. In fact, all the results of Figs.~\ref{imbalance}, \ref{spin_Delta}-\ref{Delta_orb} are plotted as functions of the electronic temperature of the confined layer. If the electron-phonon interaction is not so strong, that is $\tau_{e-ph} \lesssim \tau_G$, all our results for the order parameter are still valid, but $T_{S,N}^e$ (which is higher than $T_{S,N}$) should be substituted for $T_{S,N}$ in all the horizontal axes in Figs.~\ref{imbalance}, \ref{spin_Delta}-\ref{Delta_orb}. The opposite limit $\tau_{e-ph} \gg \tau_G$ is not of interest, because in this case the electron temperature in the superconductor is very close to the temperature of the normal reservoir $T_N$ and the thermally induced spin imbalance is negligible.

\section{Superconducting order parameter}
\label{OP}

Let us now consider how the thermally induced pure spin imbalance $\mu_S$, accumulated in the confined superconductor, influences the superconductivity there. We assume that the superconductor is in the dirty limit. The calculations are performed in the framework of the Usadel equations for the quasiclassical Green's functions \cite{golubov04,buzdin05,bergeret05,eschrig15}.
The superconducting order parameter $\Delta$ is calculated from the self-consistency equation:
\begin{equation}
\Delta = \frac{\lambda}{4}\sum \limits_{\sigma}\int \limits_{-\Omega_D}^{\Omega_D} d \varepsilon  {\rm Re}\left[F_{\sigma}\right]\tanh \left[\frac{\varepsilon-\sigma \mu_S}{2T_S}\right],
\label{selfcons}
\end{equation}
where $\lambda$ is the dimensionless coupling constant and $F_\sigma$ is the retarded Green's function for the spin subband $\sigma$.

For the problem under consideration it is enough to find the retarded Green's functions in the approximation of the impenetrable S/N interface. The reason is the following. If the junction transparency is small, as we assume, the proximity effect is of the first order with respect to transparency. But the spin imbalance is of the zero order with respect to it. As it is seen from Eq.~(\ref{spin_imbalance}), the spin imbalance is mainly determined by the temperature difference $T_N - T_S$ and the Zeeman splitting of the DOS in the superconductor. Therefore, we can safely neglect the proximity effect in calculations of the spin imbalance and the resulting order parameter. In terms of the $\theta$-parametrization \cite{belzig99} the Usadel equation for the retarded Green's function in the superconducting film takes the form:
\begin{eqnarray}
(\varepsilon \pm h)\sinh \theta_{\uparrow, \downarrow}+\Delta \cosh \theta_{\uparrow, \downarrow}+ \nonumber \\
iD\frac{e^2}{6c^2\hbar}H^2d^2 \cosh \theta_{\uparrow, \downarrow} \sinh \theta_{\uparrow, \downarrow}=0,
\label{S1}
\end{eqnarray}
The last term in Eq.~(\ref{S1}) accounts for the depairing of superconductivity in thin films by the orbital effect of the magnetic field \cite{maki}. $d$ is the thickness of the superconducting layer in the direction perpendicular to the interface plane. It is assumed to be smaller than the London penetration depth, so that the field is supposed to penetrate uniformly into the superconducting layer.

The normalized DOS can be found as
\begin{equation}
N_{\uparrow,\downarrow}={\rm Re}\left[\cosh \theta_{\uparrow, \downarrow}\right],
\label{S2}
\end{equation}
and the retarded anomalous Green's function, which enters the self-consistency equation, takes the form
\begin{equation}
F_{\uparrow, \downarrow}=\sinh \theta_{\uparrow, \downarrow}
\label{S3}.
\end{equation}

\subsection{Superconductivity recovering}

We begin by considering the case of small $T_N$, plotted in Fig.~\ref{spin_Delta}. In order to clearly see the main physical effects of $\mu_S$ on superconductivity, at first we neglect the depairing due to the orbital effect of the magnetic field. The resulting order parameter as a function of the superconductor temperature $T_S$ is represented in Fig.~\ref{spin_Delta}(a) for different magnetic fields.
The order parameter without taking into account the thermospin accumulation is plotted by dashed lines. The Zeeman depairing of superconductivity is clearly seen from these dashed curves. Solid lines represent the order parameter in the presence of the thermospin accumulation. Physically the main effect of the thermospin accumulation here is the compensation of the Zeeman depairing. It is seen from Fig.~\ref{spin_Delta} that in the presence of this accumulation $\Delta$ survives at larger temperatures as compared to dashed lines.
This is the recovering of superconductivity by creation of spin imbalance. The critical temperature increases and can sufficiently exceed its value without thermospin effect and $\Delta(T_S) \to \Delta(T_S)|_{H=0}$.

The underlying qualitative physics can be described as follows. At first let us consider the classical BCS superconductor with the gap, which is symmetric over the quasiparticle energy and spans the energy region from $-\Delta$ to $\Delta$. The symmetric location of the chemical potential at $\varepsilon=0$ provides the conditions that the number of thermal quasiparticles in the system is minimal. In our case the superconducting gaps in the two spin subbands are shifted by the Zeeman term. The same argument is valid in this case also: setting the value of the chemical potential of the corresponding subband just in the middle of the gap, we minimize the number of thermal quasiparticles. Therefore, their depairing effect on superconductivity is minimal also and it leads to the gain in the condensation energy of the system.

The dotted parts of the curves represent absolutely unstable solutions. In the regions where these unstable solutions exist, normal ($\Delta=0$) and superconducting states are stable simultaneously \cite{sarma63,larkin64,bobkova14}. Therefore, the first-order transition from the superconducting to normal state should take place at high enough magnetic fields. However, the problem of finding the exact value of the transition field is beyond the scope of the present work because we assume the spin-flip time to be the largest time scale. Nevertheless, most of the curves, represented in Fig.~\ref{spin_Delta}(a) is below the zero-temperature Pauli limiting field $h=\Delta_0/\sqrt 2$ \cite{larkin64}.

\begin{figure}[!tbh]
  \centerline{\includegraphics[clip=true,width=2.75in]{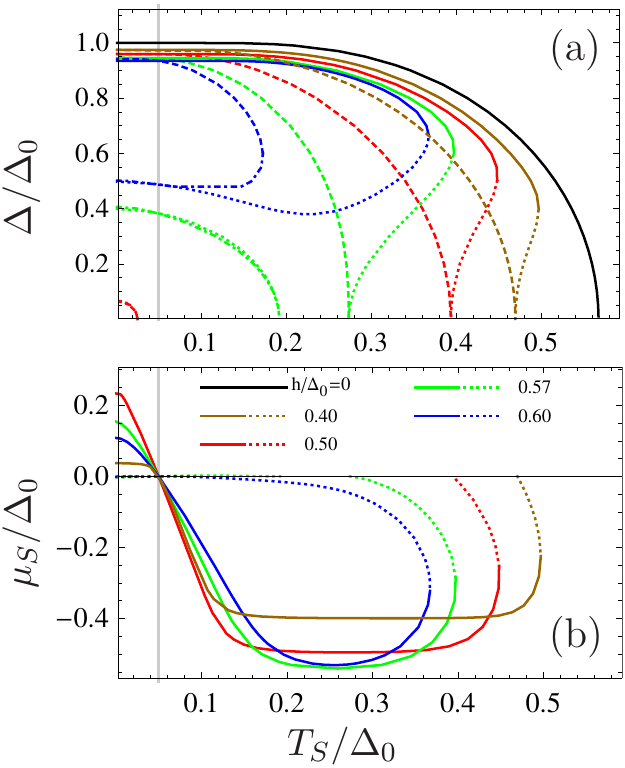}}
   \caption{(a) Superconducting order parameter for $T_N=0.05 \Delta_0$ as a function of $T_S$. (b) The appropriate spin imbalance $\mu_S$ as a function of $T_S$. The different types of lines mean the same as in Fig.~\ref{spin_Delta}. The orbital depairing parameter $\beta=0.2$.}
\label{spin_Delta_suppl}
\end{figure}

Now we take into account the realistic values of the orbital depairing by the applied magnetic field. The corresponding results are presented in Fig.~\ref{spin_Delta_suppl}.

Comparing Fig.~\ref{spin_Delta_suppl} to Fig.~\ref{spin_Delta}, we can see that the results are very similar, but now the recovering of superconductivity is incomplete at any temperatures of the superconductor. This is because now there are two different depairing factors in the system: (i) the Zeeman depairing and (ii) the depairing by the orbital effect. The Zeeman depairing can be fully compensated by the appropriate spin imbalance \cite{bobkova11,bobkova15_2}, but the orbital depairing cannot. The reason is the following. Neglecting the spin-flip processes the anomalous Green's function in the superconductor takes the form:
\begin{equation}
F_\sigma^R=\frac{\Delta}{\sqrt{\Delta^2-[\varepsilon+\sigma h +i \Gamma]}}
\label{S4}
\enspace ,
\end{equation}
where $\Gamma$ effectively takes into account the orbital depairing. Substituting this Green's function into Eq.~(\ref{selfcons}), one can see that the appropriate choice of the spin imbalance $\mu_S=-h$ fully compensates the depairing by the Zeeman term, but it does not influence the depairing by $\Gamma$ because it enters the Green's function not as an energy shift, but rather as a broadening. When the relative value of the orbital depairing grows, the smaller and smaller part of the superconductivity suppressed by the magnetic field can be recovered by the spin imbalance and the effect of recovering becomes less pronounced. The sensitivity of the system to the orbital effect of the applied magnetic field is described by the parameter $\beta=De^2\Delta_0 d^2 /(6c^2 \hbar \mu_B^2)$. Quantitatively, the relative influence of the orbital depairing with respect to the Zeeman one is controlled by the dimensionless ratio $\beta \mu_B H/\Delta_0=De^2H d^2/(6 \mu_B \hbar c^2)$ \cite{maki}, which should not be considerably larger than unity. For the typical Al parameters and $h \sim \Delta_0$ it means that $d$ should not exceed several tens of nanometers. This simultaneously satisfies the condition that the film thickness does not exceed the London penetration depth.

It is interesting to note here that the spin imbalance, which recovers superconductivity, can be induced not only by the thermal difference, but also by the spin injection from ferromagnets \cite{bobkova11,bobkova15_2}.

\begin{figure}[!tbh]
  \centerline{\includegraphics[clip=true,width=2.8in]{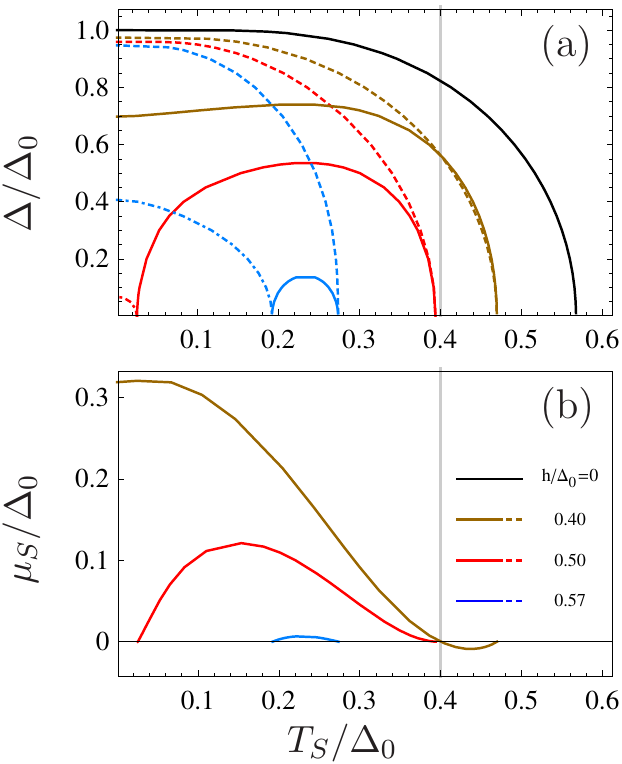}}
   \caption{(a) Superconducting order parameter for $T_N=0.4 \Delta_0$ as a function of $T_S$.  (b) $\mu_S$ as a function of $T_S$. The different types of lines mean the same as in Fig.~\ref{spin_Delta}. $\beta=0.2$.}
\label{Delta_orb}
\end{figure}

\subsection{Nonmonotonic dependence of $\Delta(T)$ and superconductivity appearance upon heating}

Now we turn to considering the case of large $T_N$, plotted in Fig.~\ref{Delta_orb}. Here the realistic orbital depairing is taken into account from the very beginning. By comparing Figs.~\ref{Delta_orb}(a) and (b) it is seen that in this case the suppression of superconductivity by the effective Zeeman field is only increased by the spin imbalance. The suppression is strongest for the lowest superconducting temperatures, where the value of the spin imbalance is maximal. In this case the superconducting order parameter shows very unusual behavior on the superconductor temperature $T_S$.  It is absent for small $T_S$ and arises upon heating the superconductor.

We can give the following physical explanation of these results. In the case when the confined superconductor is colder than the reservoir, to provide the zero spin and charge current through the interface a large number of quasiparticles is injected from the normal reservoir to the superconductor (electron-like to the spin-up subband and hole-like to the spin-down subband), see Fig.~\ref{imbalance}(c). These quasiparticles cannot annihilate each other due to the weakness of the spin-flip processes in our system and are only redistributed over energy due to the interaction with the cold phonon subsystem. As a result, we have a large number of quasiparticles in the low-temperature superconductor right above the gap. Obviously, it causes a strong depairing influence on superconductivity. At the same time, it is this quasiparticle distribution that is the spin imbalance in the system. This situation is to some extent the opposite of the well-known nonequilibrium enhancement of superconductivity predicted by Eliashberg \cite{eliashberg}. The injected quasiparticles suppress superconductivity more effectively due to the cooling by the phonon subsystem [see Fig.~\ref{imbalance}(c)].

Upon heating the superconductor it is required to inject fewer quasiparticles from the normal reservoir to provide the charge and spin balance. And also the part of the quasiparticles, located in the energy region of the coherence peak, is decreased. Therefore, the additional suppression of superconductivity by these injected quasiparticles gets lower. Under certain conditions it can result in nonmonotonic dependence of the order parameter on temperature.

 It is worth noting that the effect of superconductivity appearance upon heating is only possible for large enough magnetic fields ($h>0.5\Delta_0$), when in the absence of the thermospin effect the solution of the self-consistency equation becomes multi-valued. The point is that the system must have a possibility to reside in the normal state at $T \to 0$ in the absence of the thermospin imbalance. This possibility appears at $h>0.5 \Delta_0$, when the normal state solution becomes metastable (but not absolutely unstable) at low temperatures \cite{sarma63,larkin64,bobkova14}. In the presence of the thermospin imbalance this solution becomes the most stable.

 The effect of appearance (or at least enhancement) of the order parameter upon heating the superconductor can be viewed as a hallmark of spin imbalance in it. If the contact with a hot normal reservoir only leads to the heating of the superconductor, the nonmonotonic dependence $\Delta(T)$ is not possible. The reason is that in the absence of the spin imbalance the superconducting gap decreases monotonously with increase of the superconductor temperature.

 \section{Influence of spin-flip processes}
 \label{spin_flip}

Now we discuss the applicability of our predictions to real experimental systems. The above consideration of the effect is under the assumption $\tau_\varepsilon \ll \tau_G \ll \tau_{sf}$ because it is the most simple and clear case from the theoretical point of view. At the same time the spin-flip processes are destructive for this type of spin imbalance. In real experimental setups different regimes are possible. While for thick enough Al film very weak spin-flip rates (the spin relaxation length up to the hundreds of microns, that is the condition $\tau_\varepsilon \ll \tau_{sf}$ works very well) were reported \cite{johnson85}, the opposite case $\tau_{sf} \lesssim \tau_\varepsilon$ is realized in thin Al films (the thickness is $\sim 10$nm) \cite{hubler12,quay13,wolf13,wolf14,quay16,poli08}. But, in fact, in order to observe our predictions it is enough to realize the much weaker condition $\tau_\varepsilon \ll \tau_G \ll \tau_{sf}(\tau_{sf} \Delta_0)$.  Physically, this is because in the energy interval of interest $\Delta-h<|\varepsilon|<\Delta+h$ the elastic spin-flip processes are strongly weakened by the smallness of the DOS in one of the spin subbands. Below in order to justify this statement we estimate the influence of the finite value of the elastic spin-flip processes on the thermally induced spin imbalance.

It is convenient to turn to another representation for the distribution functions and introduce $\varphi_{0,t}=(1/2)(\varphi_\uparrow \pm \varphi_\downarrow)$. For these distribution functions the kinetic equations take the form \cite{bobkova16}:
\begin{eqnarray}
D(\kappa_1 \partial_x^2 \varphi_0 + \kappa_2 \partial_x^2 \varphi_t)-
\frac{(I_{\uparrow}+I_{\downarrow})}{2}=0, \label{S10} \\
D(\kappa_2 \partial_x^2 \varphi_0 + \kappa_1 \partial_x^2 \varphi_t)-K\varphi_t-
\frac{(I_{\uparrow}-I_{\downarrow})}{2}=0
\label{S11}
\enspace ,
\end{eqnarray}
where $\kappa_{1,2}=(\kappa_\uparrow \pm \kappa_\downarrow)/2$ and we collect the electron-electron and the electron-phonon relaxation processes into the same collision integral $I_\sigma$. $K=K_{so}+K_{mi}$ is responsible for the spin relaxation by elastic processes: spin-orbit scattering and spin-flip scattering by magnetic impurities, and \cite{bobkova16}
\begin{eqnarray}
K_{so(mi)}=16\tau_{so(mi)}^{-1}\bigl[ {\rm Re}g_\uparrow^R {\rm Re}g_\downarrow^R \mp {\rm Re}f_\uparrow^R {\rm Re}f_\downarrow^R \bigr]
\label{S13}
\enspace .
\end{eqnarray}
$K=K_{so}+K_{mi}=K(\varepsilon)$ depends on the quasiparticle energy. For the problem under consideration the most important energies are in the range $\Delta-h<|\varepsilon|<\Delta+h$ because the distribution function differs from its equilibrium value mainly in this energy interval.  Within this energy interval only the DOS for one of the spin subbands ($N_\uparrow \equiv {\rm Re}g_\uparrow^R$ for the positive energies and $N_\downarrow \equiv {\rm Re}g_\downarrow^R$ for the negative energies) is not small, as it can be seen in Figs.~\ref{imbalance}(b) and (c). The DOS for the other spin subband is zero if one neglects the spin-orbit elastic scattering and the scattering by magnetic impurities at all and of the order of $(\tau_{sf}\Delta_0)^{-1} \ll 1$ if one takes into account these self-energy terms. The same is also valid for the anomalous parts ${\rm Re}f_{\uparrow,\downarrow}^R$. Therefore for the energy interval of interest
\begin{eqnarray}
K(\varepsilon)\sim \tau_{sf}^{-1}(\tau_{sf}\Delta_0)^{-1}
\label{S14}
\enspace ,
\end{eqnarray}
that is the elastic spin-flip processes are strongly weakened by the smallness of the DOS in one of the spin subbands.

Integrating Eqs.~(\ref{S10}) and (\ref{S11}) over the width of the superconductor and making use of the boundary conditions Eq.~(\ref{S6}), one can come to the following equations:
\begin{eqnarray}
-\frac{2}{\tau_G}\Bigl[ {\rm Re}{g_0^R}(\varphi_0-\tanh \frac{\varepsilon}{2T_N})+{\rm Re}{g_t^R}\varphi_t \Bigr]=\frac{(I_{\uparrow}+I_{\downarrow})}{2}~~~~ \label{S15} \\
-\frac{2}{\tau_G}\Bigl[ {\rm Re}{g_0^R}\varphi_t+{\rm Re}{g_t^R}(\varphi_0-\tanh \frac{\varepsilon}{2T_N}) \Bigr]=  \nonumber \\
K\varphi_t + \frac{(I_{\uparrow}-I_{\downarrow})}{2}
\label{S16}
\enspace .~~~~
\end{eqnarray}
Eqs.~(\ref{S15})-(\ref{S16}), in principle, allow for exactly taking into account the spin-flip processes. But here we are only interested in qualitative estimates of their influence on the order parameter behavior in two cases: (i) the recovering of superconductivity when $T_S > T_N$ and (ii) the enhancement or appearance of superconductivity upon heating when $T_S<T_N$.

We assume that $\varphi_{0,t}=(1/2)\Bigl[ \tanh \frac{\varepsilon-\mu}{2T_S} \pm \tanh \frac{\varepsilon+\mu}{2T_S} \Bigr]+\delta \varphi_{0,t} \equiv \varphi_{0,t}^{(0)}+\delta \varphi_{0,t}$. Then we linearize Eqs.~(\ref{S15})-(\ref{S16}) with respect to $\delta \varphi_{0,t}$. The collision integrals can be written in the form $I_\sigma=I_\sigma(\varphi_{0,t}^{(0)})+\delta I_\sigma$.
For the qualitative estimates we consider the energy relaxation in the $\tau$-approximation, so that
\begin{equation}
\delta I_\sigma=\frac{\delta \varphi_0 + \sigma \delta \varphi_t}{\tau_\varepsilon}
\label{S17}
\enspace .
\end{equation}
Strictly speaking, $\tau_\varepsilon$ also depends on the spin direction, but we do not take it into account for the qualitative consideration.

Making use of Eq.~(\ref{S17}), from Eqs.~(\ref{S15}) and (\ref{S16}) one obtains the following estimates for $\delta \varphi_{0,t}$:
\begin{equation}
\delta \varphi_t=-K \tau_\varepsilon \varphi_t^{(0)}
\label{S18}
\enspace ,
\end{equation}
and
\begin{equation}
\delta \varphi_0=-2{\rm Re}{g_t^R}\frac{\tau_\varepsilon}{\tau_G} \delta \varphi_t \sim \frac{\tau_\varepsilon}{\tau_G} K\tau_\varepsilon \varphi_t^{(0)}
\label{S19}
\enspace .
\end{equation}
As far as we consider the regime $\tau_\varepsilon \ll \tau_G \ll K^{-1}$, we can neglect the correction (\ref{S19}) to $\varphi_0$, because it is much smaller than the correction (\ref{S18}) to $\varphi_t$.

Turning to the original representation we obtain that the corrections due to spin-flip processes to $\varphi_\sigma$ take the form:
\begin{equation}
\delta \varphi_{\uparrow,\downarrow}=\mp \frac{K \tau_\varepsilon}{2} \Bigl[ \tanh \frac{\varepsilon-\mu}{2T_S} - \tanh \frac{\varepsilon+\mu}{2T_S} \Bigr]
\label{S20}
\enspace .
\end{equation}
We see that under the condition $\tau_\varepsilon \ll \tau_G \ll K^{-1}$ they are small and have only minor influence on the effects we consider. In the energy interval of interest this condition is equivalent to $\tau_\varepsilon \ll \tau_G \ll \tau_{sf}(\tau_{sf}\Delta_0)$. This is much weaker than $\tau_\varepsilon \ll \tau_G \ll \tau_{sf}$ and even for the thin films used in the recent experiments \cite{hubler12,quay13,wolf13,wolf14,quay16,poli08}, where $\tau_{sf} \lesssim \tau_\varepsilon$ was reported, it is probably valid because of the smallness of the factor $(\tau_{sf}\Delta_0)^{-1}$.

At the same time $\tau_G$  can be chosen in the appropriate interval by varying the transparency of the tunnel S/N interface.

\section{Conclusions}
\label{conclusions}

In conclusion, we have shown that the giant "thermospin" effect in Zeeman-split S/N heterostructures can lead to an accumulation of a pure spin imbalance in some parts of a system. The maximal value of this imbalance is of the order of Zeeman splitting of the superconducting DOS. This thermally induced pure spin accumulation strongly influences the superconductivity in the heterostructure. Different regimes are possible: if the superconducting part is more hot than the normal one, the spin imbalance weakens the Zeeman depairing and, therefore, recovers superconductivity in the system. Otherwise, if the normal part is more hot, then the imbalance strengthens the Zeeman depairing of superconductivity. It can result in the appearance of superconductivity upon heating the sample. It is obvious that the spin imbalance also strongly influences the weak superconductivity regions, such as Josephson junction interlayers. Therefore, the effect discussed here can be used to control the appearance and positions of $0-\pi$ transitions in the S/N/S junctions via the manipulating by the temperature difference between the leads and the interlayer.

\section{Acknowledgments}

We are grateful to M. Silaev and T. Heikkil$\rm \ddot a$ for the fruitful discussions. The support by the Program of Russian Academy of Sciences “Electron spin resonance, spin-dependent electron effects and spin technologies” and
   Russian-Greek Project N 2017-14-588-0007-011 “Experimental and theoretical studies of physical properties of low-dimensional quantum nanoelectronic systems”
   is acknowledged.


\begin{thebibliography}{99}
%
\bibitem{eschrig15}
M. Eschrig, Reports on Progress in Physics {\bf 78} 104501 (2015).
%
\bibitem{buzdin05}
A.~I.~Buzdin, Rev. Mod. Phys. {\bf 77},  935 (2005).
%
\bibitem{bergeret05}
F.S. Bergeret, A.F. Volkov, and K.B. Efetov, Rev. Mod. Phys. {\bf 77}, 1321 (2005).
%
\bibitem{keizer06}
R.S. Keizer, S.T.B. Goennenwein, T.M. Klapwijk, G. Miao, G. Xiao,
and A. Gupta, Nature (London) {\bf 439}, 825 (2006).
%
\bibitem{alidoust10}
M. Alidoust, J. Linder, G. Rashedi, T. Yokoyama, and A. Sudbo, Phys. Rev. B {\bf 81}, 014512 (2010).
%
\bibitem{robinson10}
J.W.A. Robinson, J.D.S. Witt, and M.G. Blamire, Science {\bf 329}, 59 (2010).
%
\bibitem{khaire10}
T.S. Khaire, M.A. Khasawneh, W.P Pratt, and N.O. Birge,
Phys. Rev. Lett. {\bf 104}, 137002 (2010).
%
\bibitem{anwar10}
M.S. Anwar, F. Czeschka, M. Hesselberth, M. Porcu, and J. Aarts,
Phys. Rev. B {\bf 82}, 100501(R) (2010).
%
\bibitem{bergeret14}
F. S. Bergeret, I. V. Tokatly, Phys. Rev. B {\bf 89}, 134517 (2014).
%
\bibitem{gomperud15}
I. Gomperud, J. Linder, Phys. Rev. B {\bf 92}, 035416 (2015).
%
\bibitem{jacobsen15_1}
S.H. Jacobsen, J.A. Ouassou, J. Linder, Phys. Rev. B {\bf 92}, 024510 (2015).
%
\bibitem{jacobsen15_2}
S.H. Jacobsen, J.Linder, Phys. Rev. B {\bf 92}, 024501 (2015).
%
\bibitem{alidoust15}
M. Alidoust and K. Halterman, J. Phys: Cond. Matt. 27, 235301 (2015).
%
\bibitem{yang10}
H. Yang, S.-H. Yang, S. Takahashi, S. Maekawa, and S.S.P. Parkin, Nature Mater. {\bf 9}, 586 (2010).
%
\bibitem{hubler12}
F. Hubler, M.J. Wolf, D. Beckmann, and H.v. Lohneysen, Phys. Rev. Lett. {\bf 109}, 207001
(2012).
%
\bibitem{quay13}
C.H.L. Quay, D. Chevallier, C. Bena, M.~Aprili, Nature Phys. {\bf 9}, 84 (2013).
%
\bibitem{wolf13}
M.J. Wolf, F. Hubler, S. Kolenda, H.v. Lohneysen, and D. Beckmann, Phys. Rev. B {\bf 87}, 024517
(2013).
%
\bibitem{wakamura14}
T. Wakamura, N. Hasegawa, K. Ohnishi, Y. Niimi, and YoshiChika Otani, Phys. Rev. Lett. {\bf 112}, 036602 (2014).
%
\bibitem{wolf14}
M. J. Wolf, C. Surgers, G. Fischer, and D. Beckmann, Phys. Rev. B {\bf 90}, 144509 (2014).
%
\bibitem{bobkova15}
I.V. Bobkova and A.M. Bobkov, JETP Letters, {\bf 101}, 118 (2015).
%
\bibitem{silaev15}
M. Silaev, P. Virtanen, F.S. Bergeret, T.T. Heikkila, Phys. Rev. Lett. {\bf 114}, 167002 (2015).
%
\bibitem{krishtop15}
T. Krishtop, M. Houzet, J. S. Meyer, Phys. Rev. B {\bf 91}, 121407(R) (2015).
%
\bibitem{bobkova16}
I.V. Bobkova and A.M. Bobkov, Phys. Rev. B {\bf 93}, 024513 (2016).
%
\bibitem{virtanen15}
P. Virtanen, T.T. Heikkila, and F.S. Bergeret, Phys. Rev. B {\bf 93}, 014512 (2016).
%
\bibitem{quay16}
C.H.L. Quay, C. Dutreix, D. Chevallier, C. Bena, M. Aprili, Phys. Rev. B {\bf 93}, 220501 (2016).
%
\bibitem{giazotto08}
F. Giazotto and F. Taddei, Phys. Rev. B {\bf 77}, 132501 (2008).
%
\bibitem{linder15}
J. Linder, J. W. A. Robinson, Nature Phys. {\bf 11}, 307 (2015).
%
\bibitem{machon13}
P. Machon, M. Eschrig, W. Belzig, Phys. Rev. Lett. {\bf 110}, 047002 (2013).
%
\bibitem{ozaeta14}
A. Ozaeta, P. Virtanen, F.S. Bergeret, and T.T. Heikkila, Phys. Rev. Lett. {\bf 112}, 057001 (2014).
%
\bibitem{kalenkov12}
M.S. Kalenkov, A.D. Zaikin, L.S. Kuzmin, Phys. Rev. Lett. {\bf 109}, 147004 (2012).
%
\bibitem{kolenda16}
S. Kolenda, M.J. Wolf, D. Beckmann, Phys. Rev. Lett. {\bf 116}, 097001 (2016).
%
\bibitem{kolenda16_1}
S. Kolenda, P. Machon, D. Beckmann, and W. Belzig, Beilstein J. Nanotechnol. {\bf 7}, 1579 (2016).
%
\bibitem{linder16}
J. Linder, M.E. Bathen,  Phys. Rev. B {\bf 93}, 224509 (2016).
%
\bibitem{bathen16}
M.E. Bathen, J. Linder, Sci. Rep. {\bf 7}, 41409 (2017).
%
\bibitem{silaev2017}
M.A. Silaev, Phys. Rev. B {\bf 96}, 064519 (2017).
%
\bibitem{bauer12}
G. Bauer, E. Saitoh, and B.J. van Wees, Nat. Mat. {\bf 11}, 391 (2012).
%
\bibitem{slachter10}
A. Slachter, F. Bakker, J. Adam, and B.J. van Wees, Nature Phys. {\bf 6}, 879 (2010).
%
\bibitem{johnson85}
M. Johnson and R.H. Silsbee, Phys. Rev. Lett. {\bf 55}, 1790 (1985).
%
\bibitem{note1}
The thermally induced electric currents (giant thermoelectric effect) at zero voltage bias are only possible if the spin-up and spin-dow conductivities of the interface are different, for example at a superconductor/ferromagnet interface \cite{machon13,ozaeta14}. However, the giant thermoelectric effect was also reported at a S/N interface at nonzero applied voltage in the nonlinear regime \cite{kolenda16_1}.
%
\bibitem{belzig99}
W. Belzig, F.K. Wilhelm, C. Bruder, G. Schon, A.D.Zaikin, Superlattices and Microstructures, {\bf 25}, 1251 (1999).
%
\bibitem{kopnin}
N.B. Kopnin  {\it Theory of Nonequilibrium Superconductivity}, Oxford University Press, 2001.
%
\bibitem{golubov04}
A.A. Golubov, M.Yu. Kupriyanov, and E. Il?chev, Rev. Mod. Phys. {\bf 76}, 411 (2004).
%
\bibitem{sarma63}
G. Sarma, J. Phys. Chem. Solids {\bf 24}, 1029 (1963).
%
\bibitem{larkin64}
A.I. Larkin and Yu.N. Ovchinnikov, Sov. Phys. JETP {\bf 20}, 762 (1965) [Zh. Eksp. Teor. Fiz. {\bf 47}, 1136 (1964)].
%
\bibitem{bobkova14}
I.V. Bobkova and A.M. Bobkov, Phys. Rev. B {\bf 89}, 224501 (2014).
%
\bibitem{maki}
K. Maki, Gapless superconductivity. In: Parks R.D. (ed.) Superconductivity, Ch.18, p. 1035, Marcel Decker (1969).
%
\bibitem{bobkova11}
I.V. Bobkova and A.M. Bobkov, Phys. Rev. B {\bf 84}, 140508 (2011).
%
\bibitem{bobkova15_2}
I.V. Bobkova and A.M. Bobkov, JETP Lett. {\bf 101}, 407 (2015).
%
\bibitem{eliashberg}
G. M. Eliashberg, JETP Lett. {\bf 11}, 114 (1970).
%
\bibitem{poli08}
N. Poli, J.P. Morten, M. Urech, A. Brataas, D.B. Haviland, and V. Korenivski, Phys.Rev. Lett. {\bf 100}, 136601 (2008).
%





\end{thebibliography}
\end{document}